# A Visualization of the SU(2) Vacuum on the Lattice

## F. Gutbrod

Deutsches Elektronen-Synchrotron DESY
Notkestr. 85, D22603 Hamburg, Germany
e-mail: gutbrod@mail.desy.de
Homepage: www.desy.de/∼gutbrod

### Abstract

Configurations of pure SU(2) gauge field theory on the lattice are transformed to Landau gauge. After Fourier transformation, large momentum amplitudes are suppressed by a variable amount, and the configurations are transformed back to x-space. Spectacular peaks in electric and magnetic field strengths are found, which share many properties with either almost pointlike instantons in regular gauge or with extended anti-instantons in singular gauge. Environments around those peaks are visualized with respect to the action, to the topological charge density, to the gauge fields and to electric and magnetic field strengths. The density of the peaks is of order $1 fm^{-4}$, and it scales according to the string tension under a variation of the coupling constant. The evolution of the peaks under the amount of Fourier filtering is visually compared to the evolution under cooling, and the gauge dependence of the peaks is discussed. Advantages and shortcomings of this method are discussed, with emphasis on possible strong distortions of the action and the topological charge density, which become gauge dependent. Finally, I compare the character of the SU(2)-configurations to those of non-compact abelian theory.

## 1 Introduction

It is hardly necessary to justify an attempt to look directly onto gauge field configurations which have been generated by standard lattice techniques. Unfortunately, the amount of difficulty encountered in the course of such attempts is also quite obvious. On the level of physics, the freedom associated with local gauge transformations requires a careful discussion of the significance of any observations, other than those based on a few gauge invariant quantities. Even for a given gauging algorithm, the resulting configuration will depend on the application of a random gauge transformation prior to gauging, due to the Gribov ambiguities [1]. Furthermore, the short distance quantum noise degrades any visual impression. The removal of theses fluctuations will, in any case, distort the lattice configuration to an unknown degree.

On the technical level, the number of local degrees of freedom in non-abelian theories is so large that one has to strongly reduce the amount of information before something can be plotted on a screen or on paper. On the other hand, the speed of progress in computer-based visualization techniques is so high that today's obstacles may be irrelevant next year.

The present approach to dealing with the physics problems is the following: A thermalized gauge field configuration is transformed to the Landau gauge, and the gauge fields of the resulting configuration are smoothed by standard Fourier filtering procedures. Thus, the short distance fluctuations are eliminated. Presumably, these are of perturbative origin and do not deserve too much attention. To what extent this is correct, may be partly checked *a posteriori*. There are good indications that the assumption is not totally correct. These indications are connected with the observation that at a few positions on the lattice, gauging leads to very strong fluctuations of the gauge fields which look like singularities, and that at these points the Fourier amplitudes at all momenta are non-perturbatively connected. This important observation will be explained in detail in sect. 4.2, where also the nature of these special points will be described thoroughly.

In order to avoid being overwhelmed by the amount of data on a large lattice, a subset of the lattice points is selected and the associated fields are visualized on a PC-screen. The selection of what to plot and what to discard is partly made dynamically, and the techniques described in the following are motivated by the study of instantons on the lattice. After only quite modest filtering, i.e. after discarding just the highest momentum Fourier components, one can observe pronounced maxima in several quantities such as electric and magnetic fields, in the Wilson action density[1] $S^a(x)$ and in the topological charge density $P^a(x)$. Their positions are identical with those of the gauge field singularities mentioned previously. They are found numerically by a computer sweep through the lattice. After this, the selection of lattice sites is done either by considering three 2-dimensional planes through one of those maxima, or by selecting a 4-dimensional subcube (e.g. of size $9^4$ lattice unites) around the peaks.

The density of the peaks agrees with that of instantons found in the SU(2)-vacuum by cooling [2], and the visualization tools described below suggest that we are indeed seeing instantons on the lattice. Especially, the orientation and the relative strengths of electric and magnetic fields fit well with those of (anti-) instantons. However, using the action density $S(x)$ as an indicator, the size of the objects seems to be significantly smaller than the size of the instantons obtained by cooling [2]. In this connection it has to be realized that the effect of Fourier filtering on gauge configurations will be quite subtle when the gauge fields are so large or so slowly varying that the non-linear properties of field strengths are crucial. This will be discussed in sect. 2.2, where the influence of Fourier filtering on artificial lattice instantons is also studied. This influence turns out to be very strong if the instanton is realized in the singular gauge (see ref. [3] and eq. 20 for a description of this gauge). Now, it is quite difficult to decide unambiguously -on the basis of the lattice configuration- whether we are dealing with an instanton with a radius of one or two lattice units in the regular gauge, or with a broad anti-instanton in the singular gauge. This difficulty is connected with deviations of the observed gauge fields from the ideal field distribution. Thus, the visualization technique provides us primarily with reliable information on the rapid variation of the gauge fields at the special lattice points, whereas an identification of the fields with the pure case distributions will require more detailed work.

---

[1]The upper index $a$ refers to the 3 colour components of SU(2).

Returning to the description of technical tools, I list the following methods:

- A projection of all local maxima of e.g. the topological charge density onto one plane, where the maxima are defined within certain small hypercubes. A first impression is given by fig. 1, derived from a configuration on a $48^3 \times 64$ lattice at $\beta = 2.85$ which shows the presence of 5 clean peaks and one murky one with (anti-) instanton properties, plus some unqualified noise. This plot is derived via a relatively weak Fourier filtering (F.f.). For the effect of stronger filtering, see fig. 2. The details of this pictures are described in section 4.

- Gauge fields, electric and magnetic fields are plotted in three lattice planes, passing through one of those peaks.

- Field lines of electric and magnetic fields passing through the maxima are plotted for the three SU(2) colour components simultaneously. This allows one to study the quality of the wave function of the instantons as function of the distance from the center. An example can be found in fig. 3, where electric fields are plotted by 3D techniques. Rotation of the picture on the screen shows that the field lines of the three colours are orthogonal on each other, as it must be for an instanton.

- The evolution the gauge fields on a small sublattice, as a function of the cut-off, inherent in the F.f., can be visualized on the PC-screen (not on paper) by means of animation techniques. This allows to demonstrate how the maxima emerge out of the quantum noise when the cut-off in momentum space increases beyond a small fraction of an inverse lattice unit. The same animation can be performed for cooling, which demonstrates the qualitatively different response of the configuration on the two techniques. The outcome is described in sect. 5.

Further investigations include the following:

Fourier transformation requires some kind of linearization. A comparison of two different linearization techniques which are described in sect. 2.1, indicates that the phenomena described below do not depend on these technical details.

Since the gauging procedure is not unique [1], an ensemble of Landau gauges is studied by first performing several random gauges on a given configuration. This gives some insight as to what depends on the special Gribov copy into which the gauging has led, and what not (see sect. 6). A comparison with the results of other smoothing techniques, notably the cooling method[2, 6, 7], is advisable and will be attempted to a certain degree in sect. 6.2.

The interpretation of individual plots is, of course, open to subjective judgements. It has to be supplemented by measurements with sufficiently good statistics, before definite conclusions can be drawn. This criterion is perhaps not fully met in this paper, and the work has to be regarded as exploratory.

Technical information on the visualization and on the access to coloured figures is given at the end of section 3.

## 2 Fourier Filtering of Gauged Configurations

It is well known that lattice configurations are so rough that gauge invariant quantities like the Wilson action vary rapidly on the scale of one lattice unit. This has already been visualized by refs. [4] and [5] for the action and the topological density, and it holds even at the largest $\beta$ accessible to sensible Monte Carlo simulation at present. It is natural to expect that in a smooth gauge - like the Landau gauge- the gauge fields themselves share the above property. Thus, before anything can be seen in a lattice configuration, a modification of the configuration is necessary, and the ensuing distortions may be quite severe. In this paper I concentrate on Fourier filtering (F.f.) as a method of removing short range quantum fluctuation. The filtering will act on gauge fields which are obtained by linearization of lattice link variables in the Landau gauge. In this gauge, all link variables are transformed towards the continuum limit as closely as possible, with no position or direction preferred. An extension of the visualization techniques described here to non-covariant gauges is certainly possible, and it can be pursued at a later date.

It is not being claimed that this method is really superior to other smoothing techniques. However, it is conceptually quite different from the standard cooling technique [2, 6, 7, 8], and, what's more, it leads to different results. It is suggested, therefore, that it may be advantageous to have access to different types of tools when discussing in what sense something exists on the lattice.

### 2.1 Linearization and Selection of Variables

Before discussing possible advantages of the F.f., its implementation in a non-linear theory has to be defined. Naturally, in the course of this, severe ambiguities will become apparent. I will work with the representation of the SU(2) link variables

$$U_\mu(x) = u_{0,\mu}(x) + i\vec{\tau}\vec{u}_\mu(x) \tag{1}$$

with

$$u_{0,\mu}^2(x) + \vec{u}_\mu^2(x) = 1. \tag{2}$$

Before being able to use a Fourier transform, I have to define some linearization. I first transform to Landau gauge by numerically searching for a maximum of the gauge function $F(U)$,

$$F(U) = \sum_{x,\mu} u_{0,\mu}(x) \tag{3}$$

under local gauge transformations. In the weak coupling limit, all "large" components $u_{0,\mu}(x)$ should approach unity, which would easily allow linearization by working with the "small" components only. In practice, however, even at the largest values of $\beta$ used here, many link variables are far from this weak coupling limit. Especially, at $\beta = 2.7$ with the Wilson action, one finds a fraction of $5 \times 10^{-6}$ links with $u_{0,\mu}(x) < 0$. This poses a difficulty for the linearization which is necessary for a F.f.

I have used 2 different approximations for dealing with this departure from the weak coupling regime. The first one is to simply identify the (non-compact) gauge fields with the "small" components of $U_{\mu}(x)$, i.e.

$$\frac{2}{g_0} \vec{u}_{\mu}(x) \to \vec{A}_{\mu}(x). \tag{4}$$

After modifications of the $\vec{A}_{\mu}(x)$ by F.f., link variables have been reconstructed by setting $\vec{u}'_{\mu}(x) = \frac{g_0}{2} \vec{A}'_{\mu}(x)$ and

$$u'_{0,\mu}(x) = \sqrt{1 - \vec{u}'^2_{\mu}(x)}, \tag{5}$$

if $\vec{u}'^2_{\mu}(x) < 1$. Otherwise, I set $u'_{0,\mu}(x) = 0$ and normalize the $\vec{u}'_{\mu}(x)$ accordingly.

Obviously, this is a poor approximation to the link variables if $u_{0,\mu} < 0$ since an adiabatic approach to the continuum limit $u_{0,\mu} \approx 1$ should proceed first via values $u_{0,\mu} \approx 0$ and thus via increasing values of $\vec{u}^2_{\mu}(x)$. Normally, a F.f. will proceed via decreasing amplitudes almost everywhere.

In order to obtain a better mapping for the region $u_0 \approx 0$, I use a kind of stereographic projection

$$\vec{A}_{\mu} = \frac{2}{g_0} \vec{u}_{\mu} \sqrt{2/(1 + u_{0,\mu})}. \tag{6}$$

After an eventual smoothing modification of the $\vec{A}_{\mu}$ by F.f., normalized link variables can be reconstructed via the obvious inversion:

$$\vec{w} = \frac{g_0}{2} \vec{A}', \tag{7}$$

$$\vec{u}' = \sqrt{1 - \vec{w}^2/4} \; \vec{w}, \tag{8}$$

$$u'_0 = \pm\sqrt{1 - \vec{u}'^2}, \tag{9}$$

where the positive root in the last equation is taken for $\vec{w}^2 < 2$, and the negative one otherwise. This transformation brings the "south-pole" $u^0 = -1$ not to infinity, but to $\bar{w}_\mu^2 = 4$. In section 4.1 I will describe the minor differences due to the two techniques. It should be stressed again that the prescriptions are arbitrary, similar to all other smoothing techniques.

After having defined the gauge potentials $\vec{A}_\mu$, lattice Fourier transformation leads to momentum space amplitudes $\vec{A}_\mu(\hat{k})$, and these are modified under filtering by the substitution

$$\vec{A}'_\mu(\hat{k}) = exp(-\lambda^2\ \hat{k}^2)\vec{A}_\mu(\hat{k}), \tag{10}$$

where the filtering parameter $\lambda^2$ will be varied in a large region which is determined by the momenta available on the lattice. These are, for twisted boundary conditions [9] in x, y and z directions (with $\hat{k} = 2\sin\frac{k}{2}$):

$$k_{x,n} = \frac{2\pi(n+\frac{1}{2})}{L_x}, \quad n = 0,...,L_x - 1 \tag{11}$$

for colour index 2 and 3 (or green and blue), and for colour index 1 (or red) one has

$$k_{x,n} = \frac{2\pi n}{L_x}, \quad n = 0,...,L_x - 1. \tag{12}$$

Corresponding expressions hold for other directions. Thus, the minimal squared momentum is not zero but, for the large lattice of size $48^3 \times 64$, $\hat{k}_{min}^2 = 2\pi^2/L_x^2 = 0.0086/a^2$. In this case, a value of $\lambda^2 = 10\,a^2$ will eliminate almost all momentum space amplitudes, and such a cut-off is the largest one which will maintain at least some residual structures.

After transforming the fields back to x-space, electric and magnetic fields in direction $i$ are defined by open plaquettes:

$$\vec{E}_i(x) = \frac{2}{g_0}\vec{U}_{e,i} \tag{13}$$

with

$$U_{e,i}(x) = U_t(x)U_i(x+e_t)U_t^\dagger(x+e_i)U_i^\dagger(x), \tag{14}$$

and a similar expression for the magnetic fields $\vec{B}_i(x)$.

Variables to be plotted are thus the coloured $\vec{A}_\mu(x)$-fields, the electric and magnetic fields and x-space scalar products for the three colours individually: The three components of the action density,

$$S^a(x) = \frac{1}{2}(\sum_{i=1,3}(E_i^a(x))^2 + \sum_{i=1,3}(B_i^a(x))^2) \tag{15}$$

with $a=$ red, green and blue, and the three components of the pseudoscalar density,

$$P^a(x) = \sum_{i=1,3}E_i^a B_i^a(x), \tag{16}$$

which are related to the topological charge density $q(x)$ by

$$q(x) = \frac{g_0^2}{8\pi^2} \sum_{a=r,g,b} P^a(x). \tag{17}$$

The topological charge is

$$Q = \sum_x q(x), \tag{18}$$

which is integer-valued in the presence of clean (anti-)instantons. Visual results and semi-quantitative results for these variables will be presented in section 4.

## 2.2 The Pro's and Con's of Fourier Filtering

Ideally, a smoothing technique like F.f. should make significant structures in a lattice configuration somewhat broader, while reducing their amplitude more or less monotonically as a function of the cut-off parameter $\lambda^2$. At the same time, it should decrease the amplitudes of insignificant fluctuations to a large extent. This expectation may work out for the gauge fields $\vec{A}_\mu(x)$, but not necessarily for derived quantities like the field strengths and the action. The result of smoothing will also, under certain conditions, depend sensitively on the gauge even for quantities which are gauge invariant before smoothing. This holds true because the concept of momentum space gauge fields clearly is gauge dependent. Possible effects will be discussed in connection with instantons since instanton-like configurations turn out to be the most impressive objects for visualization.

Important effects of F.f. have two sources. First of all, the field strengths in non-abelian gauge theory are, at certain distances from the center of an instanton, highly sensitive to a cancellation between the abelian term $\partial_\mu A_\nu^a(x) - \partial_\nu A_\mu^a(x)$ and the nonlinear term $g_0 \epsilon_{abc} A_\mu^b(x) A_\nu^c(x)$. In the regular gauge, this cancellation occurs at large distances from the center. The effect of F.f. then is that there the action increases for increasing $\lambda^2$, whereas at small distances the opposite holds.

Secondly, the gauge fields may be so large that any linearization technique will provide us with misleading results for the field strenghts and for the action. This will happen e.g. for instantons in a singular gauge[2] or for almost pointlike instantons. Both the abelian term and the nonlinear term of the field strengths may then be distorted strongly by F.f., such that the results will be rather surprising.

Since the magnitude of such effects will depend on the instanton radius $\rho$, one has to specify $\rho$ for a quantitative discussion. According to ref. [2], we expect to see around 1.5 (anti-) instantons per $fm^4$ with an average radius of $<\rho> \approx 0.2\,fm$, which translates to $<\rho> = 4.5\,a$ at $\beta = 2.7$ and to $<\rho> = 7.1\,a$ at $\beta = 2.85$. Thus, it is interesting to study the effects of F.f. on an instanton - which has been artificially set on the lattice - with $\rho = 5\,a$.

---

[2]On the lattice, such instantons will be defined by converting the gauge fields of a regular instanton to lattice variables using the stereographic method eq. 6, by transforming the gauge fields of the singular gauge transformation (see eq. 20 below to lattice variables in the same way. Finally, this (almost singular) gauge transformation is applied to the regular gauge distribution.

In the regular gauge, the effect of a cut-off $\lambda^2 = 0.5\,a^2$ leads to a reduction of the action density at the center of the instanton by a factor 1.4. At large distances, the action increases slightly upon filtering which is contrary to expectation and which signals a possible weakness of the method. However, since the action is small in this region, the effect is not spectacular. Close to the center, the cancellation mentioned above is insignificant because the gauge fields are small, but vary rapidly. Thus, the abelian term dominates, and there the result of smoothing is as expected. Finally, we note that the distribution of the gauge fields changes quite moderately.

However, if the (artificial) instanton is realized in the singular gauge, the results after smoothing for the action and for the pseudoscalar density may be quite spectacular. Before demonstrating this, I quote the singular gauge transformation [3]. The standard instanton gauge fields in the regular gauge are

$$A_\mu^a(x) = \frac{2}{g_0}\frac{\eta_{a\mu\nu}x_\nu}{(x^2+\rho^2)},\tag{19}$$

with the 't Hooft coefficients [10] $\eta_{a\mu\nu}$ and $\overline{\eta}_{a\mu\nu}$ (used below). The singular transformation

$$U(x) = i\frac{x_\mu}{\sqrt{x^2}}\tau_\mu^+,\ \ \tau^+ = \{-i,\vec{\tau}\},\tag{20}$$

leads to gauge fields which fall off more rapidly at infinity, but which are singular at the origin:

$$A_\mu^a(x) = \frac{2}{g_0}\frac{\overline{\eta}_{a\mu\nu}x_\nu\rho^2}{x^2(x^2+\rho^2)}.\tag{21}$$

Of course, on the lattice there is neither a real singularity at short distances, nor at infinity.

Now, under the conditions mentioned above, the action develops a sharp peak, and the sign of the pseudoscalar density within the region of the peak is reversed after F.f. The rapid fluctuations of the singular gauge fields, around the center of the instanton, get weakened by F.f., but they are still clearly visible for $\lambda^2 = 0.5\,a^2$.

The sign flip can be understood easily. The gauge field distribution of a broad instanton in the singular gauge is not too different from the distribution of a regular anti-instanton distribution of radius $\rho \sim 1a$, as long as only a restricted interval of distances is considered. It is the weakening of the short range spike in the gauge fields by F.f. which converts the singular gauge instanton into a regular anti-instanton - with opposite pseudoscalar density - for a limited range of distances.

Another shortcoming of gauging is the Gribov problem [1]. On the lattice, it manifests itself in the fact that any gauging algorithm which searches for the maximum of $F(U)$, see eq. 3, will end up in one of many local maxima, at least on a large lattice. The gauge fields at these local maxima are distinct, and this affects an eventual view on the configuration, as well as average quantities like the gluon propagator. In this investigation, it has been checked to what extent certain structures are strongly gauge dependent in the sense that they appear only in one special local maximum and not in others. Before doing so, the structures will have to be investigated. They will be described in section 4, and the test is contained in section 6.1.

On the positive side of F.f., the separation into low and high momentum amplitudes seems to have principle advantages. Asymptotic freedom in non-abelian gauge theories implies that the momentum amplitudes close to the lattice cut-off can be treated perturbatively, i.e. they should be independent of possible non-perturbative configurations at low momenta. It is not clear to what extent this is correct. In section 4.2 it will be shown that the excitation of amplitudes with momenta of $O(10\ GEV)$ leads to a clustering of gauge fields around the positions of non-perturbative objects, presumably of the instantons. Thus, a possible failure of the above assertion can be studied by F.f., and this should be regarded more as an advantage than as a drawback of the method.

Other advantages of the Fourier filtering technique have to be discussed in comparison especially to the well known cooling technique, and in the investigation of special objects. Cooling drives the configuration to local minima of the action, and thus it may create objects which "exist" on the lattice for a certain number of cooling steps and disappear again in the process of further cooling. In F.f., such objects can be attributed to low momentum components. They may be masked by high frequency components and become visible only after removal of a certain amount of high momentum amplidudes, but on further smoothing they just get broader and will not disappear. This will be demonstrated in section 4.

It has to be noticed that cooling induces a more violent distortion of the lattice configuration than F.f. By applying $O(30)$ cooling steps to a configuration after gauging, the low momentum amplitudes are suppressed significantly[3], i.e. by 20% or 30%. In F.f., nonperturbative objects become visible for a cut-off $\lambda^2 \approx 0.2\,a^2$, i.e. if momentum amplitudes with $\hat{k}^2 < 0.5\,a^{-2}$ are modified by less than 10%. For such momenta, the suppression of amplitudes by cooling already amounts to a factor 3 or more.

Finally, the difference of F.f. as compared to other methods is interesting in itself. It may be that the comparison of various techniques helps to clarify in which sense non-perturbative objects exist on the lattice.

---

[3]Low momentum amplitudes are again defined by Fourier decomposition.

# 3   Visualization Techniques

The lattice variables described in the previous section are too numerous to allow a complete presentation on a PC-screen of present quality. This holds for the set of lattice points in x-space, for colour and for the Lorentz index. Therefore, I select a lattice point which is suspected to be interesting according to some criterium, predominantly because the Wilson action density has a local maximum (after F.f.) at this point. Then, certain environments around this point will be selected. These may be

- three 2D-planes (always including the t-direction) through the special point,

- a small (e.g. $9^4$ lattice units) subcube

- electric or magnetic field lines, emanating from the special point.

Another technique is to select subcubes around several (e.g. 50) maxima and project the variables into lower dimensions. Details will be given below.

   Once the variables have been selected and written to some file, a FORTRAN routine reads this file, calculates 3D-angles and colours and writes a VRML ("Virtual Reality Modeling Language", see [11]) file. In detail, each oriented variable is represented by a coloured 3D-cone, pointing into some 3D-direction which will be specified by the variable. For $S^a(x)$ and $P^a(x)$, this direction is given by the orientation in colour space, whereas for electric and magnetic fields it is given by the direction in x-space. In many cases, the SU(2)-colours are mapped onto the ordinary PC-colours red, green and blue, but colour can also be used to indicate positions in the 4th dimension. For the gauge fields $\vec{A}_\mu(x)$, the situation is not so straighforward. Whereas the SU(2)-colours will be represented as before, one of the 4 space-time directions has to be suppressed. Which ones are selected seems to be rather arbitrary, and various choices have been tried.

   The size of the cones is given by

- the sum over colour indices in the case of $S^a(x)$ and $P^a(x)$,

- or by the spatial length for $E_i^a(x)$ and $B_i^a(x)$ (SU(2)-colour $a$ fixed),

- or by the 4D-length in the case of $A_\mu^a(x)$.

   The VRML-file can be interpreted and sent to the screen by a standard browser, e.g. by the well known and free Cosmo-Player [12]. I use the freely available browser GLVIEW [13] with good results. Both tools work for Windows NT, and the search for a LINUX-browser is continuing[4]. GLVIEW allows to shift, zoom and rotate the ensemble of cones, and thus it enables a closer inspection of interesting patterns. The capacity of the browser is sufficient to plot a complete plane of a $48^3 \times 64$-lattice with 3 SU(2)-colours, e.g. for the electric fields. In this way, one easily recognizes the correlation of maxima in field strength between the three colours, as they are typical for instantons.

---

[4]There seems to be much activity in the internet in connection with a LINUX browser, but nothing definite and easily implementable is known to me.

Another important feature of VRML is animation. This allows to visualize the evolution e.g. of $P^a(x)$ as a function of increasing $\lambda^2$. For this, around 10 different values of $\lambda^2$ between $\lambda^2 = 0$ and $\lambda^2 = 0.5\,a^2$ are used to generate a sequence of small planes around a previously detected maximum of $P^a(x)$.

True 3D representations will certainly be feasible with the help of special hardware, but this will not be pursued here.

The coloured figures are enclosed in the format .png, and the figures 4 and 18, which are black/white, are enclosed as .ps-files. The reader is also referred to my homepage www.desy.de/~gutbrod, where, in the link "publications/visualization.html", one can study and print the figures in the .jpg format, albeit in somewhat lower quality.

# 4 Results of Visualization

## 4.1 Global Plots

I will start the presentation of selected examples by an overview over full lattice configurations, where local maxima of the action density have been used to identify "hot spots". In detail, the lattice of size $48^3 \times 64$ is subdivided into sublattices of size $8^4$, and positions of the local maxima of the Wilson action are determined within these sublattices. This is done after Fourier filtering with a cut-off in the range $\lambda^2 = O(1a^2)$. If a pair of maxima has a distance less than $3\,a$, the lower maximum is discarded. The lattice positions of an environment around the maxima are stored, as well as the corresponding values of the pseudoscalar density $P^a(x)$ (without colour summation). In fig. 1, $P^a(x)$ is plotted, with 3 space-time dimensions properly respected. The fourth dimension, $z$, is converted to colours interpolating between red, corresponding to $z = 1$, and green, corresponding to $z = L_z = 48$. The $y-$direction becomes visible properly only when the plot is rotated around the (vertical) $x-$axis. This is accomplished with the help of the VRML-browser. The direction of the cones on the plot is given by the direction of $P^a(x)$ in colour space.

The configuration belongs to an ensemble of 10 thermalized configurations at $\beta = 2.85$, with twisted boundary conditions [9]. The cut-off parameter $\lambda^2$ has been chosen as $\lambda^2 = 0.3\,a^2$. In fig. 2 the same quantities are plotted, but with $\lambda^2 = 1.0\,a^2$. The overall character of the plots is the same for all configurations. Intensive investigations with different lattices sizes, with values of $\beta$ in the range $2.5 \leq \beta \leq 2.85$ and with random gauge transformations before the Landau gauging procedure have shown that the character of these plots is by no means accidental.

The surprising structure of the pictures, especially that of fig. 2, is the clear separation between six pronounced spikes and a noisy background, without any objects with interpolating character. Four spikes are nicely parallel to the diagonal in colour space, and two are antiparallel. Within the spikes, the action density equals the pseudoscalar density within 20% (this is not shown). More detailed investigations, which will be explained in the next section, show that the spikes share many properties of (anti-) instantons. Surprising facts are their small radii and an early corruption of the instanton wave functions as a function of distance from the center of the peak. Anticipating the outcome of the discussions given below, it is more likely that the objects are broad instantons in a singular gauge than almost pointlike instantons in a regular gauge.

Only minor details of the plots depend upon which of the two methods of linearization, described in section 2.1, was used. Especially within the spikes, the pseudoscalar density may differ by 10% between the two methods, and by much less outside the spikes.

The suppression of a transition region between the spikes and the background can be made more quantitative, and, in connection with this, the density of the spikes and its scaling behaviour can be determined. On each of the ten configurations, I have identified $O(1.500)$ local maxima of the action density. A value of $\lambda^2 = 0.5\,a^2$ has been used. Then, the action has been summed within a radius of $R = 2a$ around the positions of the maxima which are called $x_m$. The resulting contribution to the action will be denoted by

$$\Sigma_m(\lambda^2, R) = \sum_x^{(x-x_m)^2 \leq R^2} S(x).$$  (22)

The number of $\Sigma_m(\lambda^2, R)$-values found in intervals of unit length, summed over all 10 configurations, is shown in fig. 4 as a histogram with full lines. Because of the clear separation of the peak in the histogram at high values from the background, the number of spikes per configuration has a relatively small systematic error. One finds $6.1 \pm 0.8$ spikes per configuration, which transforms to a density in physical units

$$D_{sp}(\beta = 2.85) = 1.4 \pm 0.3\,fm^{-4}$$  (23)

if the value of $a(\beta = 2.85) = 0.028\,fm$ of ref. [14] is used. This density is consistent with the number of instantons given in [2] at smaller values of $\beta$, which has been obtained by the cooling method[5]. A similar density has been found for SU(3) in ref. [6], but it is smaller by a factor five than the result for SU(3) obtained in [7] by somewhat stronger cooling. It has to be mentioned that predictions for the instanton density play an important rôle in the search for non-perturbative phenomena in deep inelastic scattering [15, 16]. Before the present result with its relatively small error can be used in this context, a better understanding of the nature of the peaks is mandatory.

---

[5]This agreement does not extend to the observed radii. As will be discussed below, the spikes seen in F.f. have a radius of one or two lattice units, whereas in [2] an average radius of 8 lattice units has been found.

Also included in fig. 4 is the summed action $\Sigma(\lambda^2, R)$ for 20 configurations on a lattice of size $32^3 \times 64$ at $\beta = 2.7$. This histogram is marked by a broken line. The corresponding plot of one configuration with spikes is shown in fig. 5. It is obvious that the character of this configuration is slightly different from the one shown in fig. 2. At the lower value of $\beta$, a transition region between pronounced peaks and the background is clearly visible, and the same can be observed in the histogram.

For a scaling test of the density of peaks, the vertical scale in the histogram for $\beta = 2.7$ has been scaled by the ratios of the volume (in physical units) in comparison to the case $\beta = 2.85$, i.e. by

$$\frac{V_{phys}(\beta = 2.85)}{V_{phys}(\beta = 2.7)} = \frac{V_{latt}(\beta = 2.85) \, a^4(\beta = 2.7)}{V_{latt}(\beta = 2.7) \, a^4(\beta = 2.85)}. \tag{24}$$

The size of the lattice constant $a(\beta = 2.7)$ has been taken from ref. [17]. This leads approximately to the same area under the spike region. More quantitatively, at $\beta = 2.7$ there are $13.4 \pm 2$ spikes in a configuration, and we find

$$D_{sp}(\beta = 2.7) = 1.6 \pm 0.4 \, fm^{-4}. \tag{25}$$

If one derives the ratio of the scales from the density of spikes per configuration at $\beta = 2.7$ and $\beta = 2.85$, one obtains

$$a(2.7)/a(2.85) = 1.64 \pm 0.08, \tag{26}$$

which is in very good agreement with the result from the string tension [14], i.e. $a(2.7)/a(2.85) = 1.60 \pm 0.06$.

Scaling of the density alone does not necessarily imply that we are dealing with physical objects. If the spikes were instantons in a regular gauge with a radius of a couple of lattice units (assume, for definitenes, a radius $\rho = 5 \, a$ at $\beta = 2.7$, corresponding to a value $\rho = 0.22 \, fm$), then at $\beta = 2.85$ the radius should have increased at least by a factor 1.5, i.e. to $7.5 \, a$. Since we sum the action only within a radius of $2 \, a$, the quantity $\Sigma(\lambda^2, R)$ would essentially be proportional to the action density of the instanton around the center. This, however, decreases as $\rho^4/(x^2 + \rho^2)^4$, i.e. rather strongly with $\rho$ for $x^2 \ll \rho^2$. For a scaling behaviour of the radius, the peak of the histogram should have been shifted to the left by a factor 4, when going from $\beta = 2.7$ to $\beta = 2.85$. Within our statistics, the peak has moved to the right. This excludes scaling of the radius, and there is good evidence that the objects have a pointlike character. It has to be remembered, however (see section 2.2), that extended artificial instantons in a singular gauge behave abnormally under F.f., as the action develops an unphysical spike, and, within this spike, the pseudoscalar density reverses sign. At present, I see no convincing method to resolve this ambiguity. A safe conclusion is that in the Landau gauge the short range spatial variations of the gauge fields cannot be described fully by perturbation theory. Some lattice points, with a density of $\sim 1.4/fm^4$, show a violent oscillation even after modest F.f., which leads to significant peaks in the field strengths, presumably via the term $\partial_\mu A_\nu^a - \partial_\nu A_\mu^a$, without compensation by the non-linear term. The detailed properties of these variations will be studied in the next subsection.

## 4.2   Properties of the Spikes

The visualization technique can be used to study how well the spikes, observed in the global plots, share properties of instantons. These properties are

1. a strong localization of the action density in space-time,

2. selfduality, i.e. $E_i^a = \pm B_i^a$, $a = r, g, b$, $i = x, y, z$,

3. colour democracy, i.e. all colours are excited with equal strength, and electric (magnetic) fields of different colours are mutual orthogonal,

4. and a special dependence of the gauge fields on the distance from the center, which leads to a cancellation between the linear and the non-linear terms in the electric and magnetic fields. In the regular gauge, this occours at large distances.

In fig. 6, I show the pseudoscalar density on a lattice plane running through one of the spikes of fig. 2, with a cut-off $\lambda^2 = 0.15\,a^2$. It is impressive to note the strong localization and the vastly different level of the spike amplitude, as compared to the background intensity. It is also obvious that F.f., with the relatively low value of the cut-off, does not really smoothen the background, which is still quite noisy. Primarily, the F.f. reduces the background intensity such that it does not mask the peak. In addition to the strong localization, the plot also shows that the direction of the spike is parallel to the diagonal in colour space. This is consistent with colour democracy.

In fig. 7, the gauge field distribution in the same plane as above is shown for the red colour, where the fourth space-time component is disregarded in calculating the direction of the cones[6]. At the position of the spike, one notices a slight increase in the activity of the gauge fields, which seems to be statistically barely significant. But, checking for the other colours and for more configurations, the same phenomenon always shows up, and the correlation between the magnitude of gauge fields and the spike in the action has to be accepted as real.

The observation of a peak in the gauge fields allows to ask how well components of high momentum are determined by perturbation theory. For a cut-off $\lambda^2 = 0.5\,a^2$, the peak in the gauge field intensity is reduced by about a factor

---

[6]The size of the cones, however, is proportional to $(\sum_{\mu=0,3}(A_\mu^{red})^2)^{1/2}$.

2. This has important consequences. It implies, that those Fourier components which get removed when proceeding from $\lambda^2 = 0.15\,a^2$ to $\lambda^2 = 0.5\,a^2$, are coupled dynamically to the non-perturbative object which is responsible for the spike. In momentum space, these components have a momentum $k^2 \approx 2/a^2 \approx 120\,GeV^2/c^2$. Obviously, as already mentionend in section 2.2, Fourier amplitudes around these momenta cannot be determined completely by perturbation theory.

The quality of the wave function of the hypothetical instantons can be studied by plotting e.g. electric field strengths. In fig. 8, I show the coloured electric field strengths, attached to lattice points in the same plane as above, but with a smaller environment around the spike selected. The cut-off is again $\lambda^2 = 0.50\,a^2$. One nicely observes how the maxima of absolute values pop up in the same region for all colours, and how the field strengths of different colours are mutually orthogonal. It is evident, however, that there are several other locations where the field strengths are non-negligible. The amplitudes of these excitations amount to $1/2$ or $1/3$ of the maximal values at the spikes. The nature of this background has to be explored more thoroughly in the future.

In the close vicinity of the spike, one notices that in the upward direction the red field distribution has a long tail. For the green colour, one observes a flip in the orientation at a distance $\Delta$ from the center of about 3 lattice units. Furthermore, the blue fields flip sign when moving right. Thus, a possible instanton wave function is strongly distorted for $\Delta \geq 3a$, if it is realized in the regular gauge[7]

It has to be tested whether an anti-instanton realized in the singular gauge provides a better description. Consequently, in fig. 9 I show the fields of a broad artificial instanton with $\rho = 15\,a$ in singular gauge, subject to the same F.f. The field lines are approximately parallel straight lines which flip orientation at a distance of $\Delta = 5a$. This comes as an interplay of the singular gauge and the cut-off[8]. This flip is also noticable in the previous figure for some colours and some directions. A scan through all configurations and directions reveals that this phenomenon is frequent. It is one of the indications that we are dealing with singular gauges and less likely with pointlike instantons. On the other hand, the long tail of the field with red colour observed in the previous figure, is also not reproduced by this artificial anti-instanton.

In fig. 3, I display field lines. Starting from a local maximum of $S(x)$, electric field lines for all colours have been drawn, where the field strengths off the lattice sites have been determined by interpolation. When rotating this picture with the help of the VRML-browser, on the one hand, one clearly checks for the orthogonality of the field lines close to the point of intersection, but, on the other hand, one realizes that after 4 cones (two lattice units) the field lines depart from the original direction. Again, this means that the wave function of the object no longer agrees with that of an instanton. A plot for the magnetic field lines shows exactly the same trends, consistent with selfduality.

---

[7]Significant fluctuations at a distance of less than 3 lattice units are not expected to occur, since the cut-off suppresses Fourier strongly amplitudes with $k^2 > 1/a^2$.

[8]For a stronger cut-off, the field lines become more parallel.

Proceeding to larger values of the cut-off, i.e. $\lambda^2 = 4\,a^2$, I show, in fig. 10, the electric fields of red colour in a plane through the peak. Now, deviations from the instanton configuration (parallel field lines) occur at a distance $\Delta$ of 10 $a$, and at a distance $\Delta = 15\,a$ one can observe a few other, less pronounced maxima. This characteristic distance has to be interpreted in connection with the cut-off in momentum space. If we take the condition $k\Delta = \pi$ for the occurence of a significant deviation and require $\lambda^2 k^2 < 0.5$, we obtain $\Delta > 8.9\,a$, in agreement with the visual impression. In other words, the wave function is disturbed as soon as allowed by the cut-off.

Next, in figs. 11 to 13 the gauge field distributions in the same plane as previously are shown, again for $\lambda^2 = 4\,a^2$. Now, the peaks in the action and in $P^a(x)$ are clearly correlated in space to a multi-dimensional zero in the gauge fields, i.e. a zero for all directions and for all colours. This zero has evolved in quality when coming from smaller values of the cut-off, where it is obscured by the singular nature of the unsmoothed gauge fields.

We note that the zero is coupled to regions of large gauge fields. However, there are several other excitations of similar magnitude visible. Furthermore, the environment of the instanton does not seem to be the only region where rapid variations of the gauge fields occur. Among others, one notices the occurence of several vortex-like objects, some of those within a distance $\Delta < 15\,a$. The correlation of the action density with these objects remains to be investigated.

Whereas the spikes are phenomena with a well defined density and, apart from the radial dependence, with clear properties of instantons, the maxima of the "background" are more difficult to interpret. From fig. 4 we conclude that they are much more numerous than the spikes, and that they do not obey scaling of the radius, by the same arguments as given for the spikes. To isolate possible candidates of instantons with larger radii will require different filtering techniques as have been used up to now.

## 5    Cooling vs. Fourier Filtering

In this section, I study the connection between the peaks emerging after Fourier filtering, and peaks emerging after cooling. For both smoothing techniques, I start from the same configuration in the Landau gauge. There are two possibilities for a comparison. Either, one observes the gauge fields after cooling around those lattice points which are selected in the search for maxima in F.f., or one searches for maxima in the cooled configuration independently and plots the environment around those positions.

Before discussing the outcome in detail, the cooling technique has to be specified. There one replaces the link variables iteratively by a weighted sum of the link variables $U_\mu(x)$ and the staples $\Sigma_\mu(x)$,

$$U_\mu'(x) = \mathcal{P}((1-c)U_\mu(x) + \frac{c}{6}\Sigma_\mu(x)), \tag{27}$$

where $\mathcal{P}$ stands for projection onto SU(2)-matrices. The links are updated following an arbitrary path through the lattice. Based on comparison with renormalization group mapping, a value of $c \approx 0.5$ is recommended [2].

It is obvious that this method, after sufficiently many iterations, will drive the lattice configuration to a superposition of local action maxima[9] which are determined by the local boundary conditions (generated by concurrent maxima), thereby minimizing the global action. These maxima may be instantons. The density of instantons and anti-instantons, generated in this way, is a strongly decreasing function of the number of cooling steps [2, 7], whereas the difference between the number of instantons and anti-instantons depends not so strongly on cooling. This allows to determine the phenomenologically interesting quantity $<Q^2>$ within reasonable errors.

To start a comparison, I refer to the global plot of the pseudoscalar density $P^a(x)$ of the configuration used in figs. 1 and 2 on the large lattice at $\beta = 2.85$. There, the positions of the highest peaks, obtained after F.f., had been selected, and the cut-off was $\lambda^2 = 0.3\,a^2$ and $\lambda^2 = 1.0\,a^2$ resp. In fig. 14 the same configuration is shown, after 30 cooling sweeps with $c = 1$ (see eq. 27). The cooling started from Landau gauge[10], and the maxima were selected after cooling. The scale of the cones has been increased by a factor 2.5 as compared to fig. 2.

Obviously, the overall characters of the plots, obtained by F.f. and by cooling, are quite different. In the latter case, the peaks are much broader and have smaller amplitudes at their centers. The alignment along the diagonal in colour space is much more pronounced, and there is only one prominent peak as compared to the six objects visible in fig. 2 (remember the different scale in the two figures!).

From these plots, a direct comparison between peaks, eventually located at the same positions, is not easy. In order to compare corresponding peaks in the two cases, it is necessary to specify the positions of the peaks obtained by F.f. and to plot $P^a(x)$, after cooling, at these locations. The outcome of this procedure is the following: One again finds maxima around the positions of the peaks, apart from a shift of the location of the maxima by a few lattice units. The spatial extension of the peaks is much larger, their amplitudes are reduced by a factor $\approx 100$, and, most importantly, the sign of $P^a(x)$ has changed. The origin of this behaviour is clearly demonstrated in fig. 15, where the gauge fields with red colour are displayed[11] around the position of the peaks found in F.f. Obviously, by cooling the configuration evolves towards a gauge singularity, where the gauge fields are maximal in absolute size and vary most rapidly (changing sign) at the position of the peaks.

---

[9]The term maxima refers to local spatial maxima, and cooling brings the configuration to global action minima.

[10]This is irrelevant for the discussion of the variable $P^a(x)$, but not for the comparison of the gauge fields. The gauge independence of the action and of the pseudoscalar density will be exploited in the next section.

It is instructive to monitor the influence of cooling and F.f. by an animated presentation[12]. After each cooling step and after each increase of $\lambda^2$ by $\delta\lambda^2 = 0.05\,a^2$, the corresponding gauge fields and $P^a(x)$ around a selected peak are written to a file. The VRML language allows to convert the cooling/filtering steps into a time evolution. Depending on the number of sites in the plot, the change of the gauge fields in time is reasonably smooth, due to the linear interpolation offered by VRML. One observes that the gauge fields far away (by 3 lattice units or more) from the site of the peak decrease, in both schemes, more or less in the same way. However, the gauge fields which are located closely to the site remain in size essentially unchanged under cooling, with a few link variables rotating through colour space. This leads to the singularity in the gauge fields. As an effect of the minimization, the action and $P^a(x)$ are quite small. This comes about via the cancellation of the linear and the quadratic terms in the field strengths[13]. In F.f., on the other hand, the decrease in amplitude is uniform, and the gauge fields develop their smooth zero at the point of the spike. Now, the action and $P^a(x)$ are not small, since the typical cancellation mentioned before is not operative.

It has already been mentioned that far away from the spike, i.e. for $\Delta > 3\,a$, the gauge fields are similar in both schemes (in detail, the fields are somewhat more rough after cooling than in F.f., and their absolute magnitude depends on the amount of cooling and, in the case of F.f., on $\lambda^2$). Thus, one can say that the two schemes lead to a different interpolation between regions of blocked gauge fields. In F.f., this connection is a smooth one with a zero at the peak position, leading to an approximate instanton (or anti-instanton) wave function with small radius $\rho$ in the regular gauge. After cooling, the connection is achieved by a singularity, leading to the wave function of a broad anti-instanton (instanton) in the singular gauge. Probably, it is very difficult to decide which scenario is the "correct" one. In principle, the behaviour of the gauge fields at $\Delta \approx 1a$ and at $\Delta \geq \rho$ should provide an unambiguous tool for distinguishing between the two cases. However, in the first region, lattice artifacts prevail, and in the second region the wave functions are distorted heavily after a couple of lattice units, as can be seen in fig. 3.

Finally I note that the outcome of cooling is influenced by high momentum amplitudes. If one applies, before cooling, a F.f. with a small value of the cut-off, $\lambda^2 = 0.2\,a^2$, the configuration evolves to a smooth field distribution when being cooled.

---

[11]It has to be emphasised that the colour in this representation indicates the position in the fourth dimension.

[12]Unfortunately, it seems to be impossible to demonstrate this in a publication.

[13]This can be easily demonstrated by retaining only the linear term in the fields, in which case one observes the same spikes as in F.f.

In summary, both cooling and F.f., when applied to a configuration in Landau gauge, lead to a set of points (with a density of 1.4 points per $fm^4$) where the gauge fields in the environment are connected by a rapid flip of signs for all amplitudes. These points signal the impossibility to find a gauge which is smooth, after the removal of perturbative fluctuations. When one does the latter with the help of cooling, the connection between regions of different sign is achieved by a singularity in the gauge fields, and this allows for a smooth and minimal distribution of the action. After F.f., the gauge fields are interpolated smoothly (details depend on the cut-off), and this leads to a spike in the action. The latter fact is due to the inadequacy of linear modifications at the singular point, such that the subtle cancellation does not work. Anyhow, for both methods, visualization is a good tool to study the evolution of variables as a function of the number of cooling steps or of the amount of Fourier filtering.

# 6 Gauging Ambiguity

Gauging is an essential tool in the present investigation, and a natural reservation is based on the corresponding ambiguities. The latter are given by the type of gauge (Landau gauge vs. Coulomb, axial or temporal gauge) and by the Gribov ambiguities [1] within a particular gauging scheme.

## 6.1 Gribov Ambiguities

In order to study the influence of the Gribov ambiguities, five configurations at $\beta = 2.7$ have been 12 times subject to the following procedure: Before transforming to the Landau gauge, a random gauge transformation has been applied to the configuration. In about 20% of these 12 trials, the ensuing Landau gauge fixing algorithm ended up in one specific gauge. For each gauge, local maxima of the action density have been determined in the same way as described in section 4, and the positions of the 15 strongest maxima have been compared. A shift of 2 lattice units in every direction was allowed for considering the positions as identical. For all 5 configurations, there is at least one spike which occurs at the same position in all of the 12 gauges. For about 20% of the maxima, one gauge missed one of the spikes, and for 2 misses, there is a similar probability.

It has to be concluded, that the positions of the spikes are not completely unambiguous properties of the gauge orbit of a given lattice configuration. But they are also not fully a gauge artefact either. Apparently, the configuration has, at certain positions, a distortion which acts as an obstruction against the interpolation by a smooth gauge. The gauging algorithm either may find a gauge field realization which leads to strong gradients at these positions, or a smooth one, shifting the problem to another region of space-time.

It is interesting to note that in the Coulomb gauge the existence of gauge singularities is tantamount to the occurrence of Gribov ambiguities [18]. Since we can choose the coupling constant $g_0$ small enough such that no spikes are present in our configurations, it will be possible to test the connection between the phenomena numerically, albeit in Landau gauge.

## 6.2   Spike Positions in F.f. and in Cooling

The gauge independence of the action and of the pseudoscalar density under cooling offers another possibility to study the gauge dependence of the F.f. method. For several configurations, I have applied 15 cooling steps with $c = 0.5$, and afterwards the pseudoscalar density $P^a(x)$ has been plotted on planes running through spikes found under F.f. These plots can be compared directly with those obtained by F.f., e.g. with a cut-off $\lambda^2 = 1.0\, a^2$. In almost all cases it turns out that at the position of the spike there exists a maximum of $\mid P^a(x) \mid$ in the cooled configuration. Furthermore, a strong correlation exists also between most of the secondary peaks in F.f. and in cooling. The spatial extension of these peaks as well as the orientation of $P^a(x)$ in colour space differs in the two cases. This is not surprising. After all, there is a strong dependence of these quantities both on the number of cooling steps and on the cut-off used in F.f., and different quantities are being optimized in the course of the distortion of the lattice configuration.

One can conclude that the gauge dependent peaks obtained by F.f. show up at essentially the same positions as those emerging from the gauge independent cooling method. This makes it quite unlikely that in F.f. we are dealing with gauge artefacts. However, the physical properties of the peaks are necessarily different in the two methods.

# 7   Comparison with QED

How can the vacuum of SU(2) lattice gauge theory be distinguished from the free field case, i.e. from non-compact QED? The latter will simply be defined by a superposition of plane waves on the lattice. After creating such configurations, they will be subject to the same Fourier filtering as the SU(2)-vacuum, without encountering the problems of linearization, of course. For QED, no gauge singularities are expected and none are found. Thus, both vacua differ in a fundamental aspect. However, if a lattice plane is selected which does not pass through a spike, it is impossible to differentiate between both theories by inspecting plots of the action or of the gauge fields. An example is given by a comparison of fig. 16 with fig. 17, for a cut-off $\lambda^2 = 2\, a^2$ and a lattice size of $48^3 \times 64$. For SU(2), $\beta = 2.85$. The author has not been able to pin down a significant difference in the size of the gauge field clusters, which survive the smoothing procedure.

In order to make the comparison more quantitative, the correlation of gauge fields around the various maxima is studied. For a fixed colour $a$, the gauge fields around many (numerically determined) maxima are averaged within a distance of 2 lattice units. This average will be denoted by $\overline{A}^a_\mu(x)$. The same average is taken around a center shifted in the direction of the local field by a distance $R$. The scalar product of the two averages is taken, and an effective longitudinal mass $m_l(R)$ is determined by

$$m_l(R) = -\ln \sum_\mu \overline{A}^a_\mu(x)\overline{A}^a_\mu(x + e_A R)/\sum_\mu (\overline{A}^a_\mu(x))^2, \tag{28}$$

where $e_A$ is a unit vector into the direction of $\overline{A}^a_\mu(x)$. Because we move around a maximum of the square of the gauge fields, the slope at small values of $R$ is small. In a similar way, transverse correlations are defined for the gauge fields. In fig. 18, the transverse and longitudinal masses both for QED and for SU(2) are compared. The lattice size is $32^3 \times 64$, and for SU(2) I take $\beta = 2.7$. In agreement with expectation, the masses in SU(2) are significantly smaller than in the case of QED. The reason for this phenomenon lies in the behaviour of the gluon propagator in Landau gauge [21, 22] which increases more strongly towards small momenta than the free propagator $G_0(q^2) = 1/q^2$. Transformed to x-space, this implies a softer decrease of the correlation in SU(2) as compared to QED. Although the differences in the correlation are significant, they are too small as to be visible in a single configuration.

Not so clear is the behaviour of the correlations at large $R$. Especially in the transverse case, the correlation in QED seems to develop a zero around $R = 7a$, which is, of course, cut-off dependent. In SU(2), no zero is found. An explanation of this phenomenon is outside the scope of this paper.

Based on the picture of stochastically oriented domains of gauge fields [19, 20], I had expected to obtain a rapidly vanishing correlation in the non-abelian case, i.e. an increase of the effective masses as functions of $R$. This has not yet been observed. Presumably, the distances $R$ used here are too small as to see this effect.

## 8   Conclusions

There is no doubt that computer (PC-) technique has matured sufficiently such that an instructive visualization of large sublattices in SU(2) lattice gauge theory is feasible. Still, the huge number of variables imposes some restrictions in the presentation of lattice variables, but these are not so strong as to render the information obtained by visualization meaningless. This is demonstrated by the observations described in the previous sections, which were new at least to the author. In recapitulation, I found

1. the existence -in Landau gauge- of objects with a rapid variation of gauge fields which persist through Fourier filtering up to large values of the cut-off. Under this technique, they show up as narrow peaks in the (gauge non-invariant) action and pseudoscalar density. The density of these objects is $1.4\,fm^{-4}$, it scales according to the string tension, and it agrees with the density of instantons found by a special cooling method [2].

2. The properties of these objects, e.g. the orientation of electric and magnetic field lines within the peaks, both in $x$-space and in colour space, are those of instantons.

3. The peaks are well separated from a background which does not resemble the characteristics of instantons in any obvious way, at least according to first impression. More work is necessary here.

4. The strong concentration of field strenghts in $x$-space is compatible either with the existence of very narrow ($\rho \approx 1a$) instantons in regular gauge or with the existence of broad instantons ($\rho > 10\ a$) in singular gauge. The latter case is favoured by the fall-off of the gauge fields as a function of the distance from the center.

5. The quality of the wave function of the instantons, as measured by the field strengths, is strongly corrupted for distances from the center, which are smaller than the radius $< \rho >$, quoted in the literature.

6. The different response of the gauge field configuration and of other variables either to Fourier filtering or to cooling can be demonstrated impressively by animation.

7. The positions of the spikes are partly gauge dependent, i.e. some differ for different Gribov copies. However, all spikes are associated with action maxima which appear after cooling, some of which have quite small values of the action close to the center.

8. The similarity of QED and SU(2) in their respective gauge field distributions, apart from the existence of the gauge singularities, can be made visible.

It is to be expected that with more experience additional insight into other structures of the vacuum will be obtained.

## Acknowledgement

The author is indebted to R.L. Stuller and to H. Joos for useful discussions and for encouragement. The preparation of the SU(2)-configurations on the large lattice has been accomplished on 108 nodes of the T3E parallel computer at the NIC at the Forschungszentrum Jülich. The author is grateful for granting computer time and support, especially to H. Attig. The development of the visualization tools has been carried out on a dedicated SNI Celsius workstation. The author is indebted to the DESY Directorium for providing access to this facility.

# References

[1] S. Gribov, Nucl. Phys. **B139** (1978) 1

[2] Th. Degrand, A. Hasenfratz and T. Kovács, Nucl. Phys. **B 520** (1998) 301

[3] T. Schäfer and E.V. Shuryak, Rev. Modern Phys. **70** (1998) 323

[4] M.C. Chu, J.M. Grandy, S. Huang and J.W. Negele, Phys. Rev. **D49** (1994) 6039

[5] C. Michael and P. Spencer, Phys. Rev. **D50** (1995) 4691

[6] A. Hasenfratz and C. Nieter, Physics. Lett. **B 439** (1998) 366

[7] UKQCD Collaboration, D.A. Smith and M. Teper, Phys. Rev. **D58** (1998) 104505

[8] M. Albanese et al., Physics Lett. **B192** (1987) 163

[9] D. Daniel, A. Gonzáles-Arroyo and C.P. Korthals Altes, Phys. Lett. **B251** (1990), 559, and references quoted therein.

[10] G. 't Hooft, Phys. Rev. **D14** (1976) 3432

[11] A.L. Ames, D.R. Nadeau and J.L. Moreland, The VRML Source Book, 2nd ed., J. Wiley & Sons, New York 1997

[12] Cosmo-Player, download from www.cai.com/cosmo/

[13] H. Grahn, download of GLVIEW 4.3 from www.snafu.de/~hg

[14] S.P. Booth, A. Hulsebos, A.C. Irving, A. McKerrel, C. Michael, P.S. Spencer and P.W. Stephenson, Nucl. Phys. **B394** (1993), 509

[15] A. Ringwald and F. Schrempp, Nucl. Phys. Proc. Suppl. **79** (1999) 447

[16] A. Ringwald and F. Schrempp, Physics Lett. **B459** (1999) 249

[17] A. Hulsebos, Nucl. Phys. **B30** (Proc. Suppl.) (1993) 539

[18] R.L. Stuller, Phys. Rev. **D22** [1980] 2510

[19] H.G. Dosch and Yu.A. Simonov, Physics Lett. **B 205** (1988) 339

[20] P. v. Baal, Nucl. Phys. Proc. Suppl **63** (1998) 126

[21] F. Gutbrod, DESY 96-252 (December 1996) (unpublished)

[22] UKQCD, D.B. Leinweber, J.I. Skullerud and A. G. Williams and C. Parrinello, Phys. Rev. **D60** (1999) 031501

## Figure Captions

The figures (mostly coloured) are in the format .png, except the two black-white ones, which are in the .ps format.

Figure 1: Large lattice ($48^3 \times 64$, $\beta = 2.85$): Pseudoscalar density maxima, projected into 3 dimensions. Colour indicates position of maximum in fourth dimension, moving from red (z = 1) to green (z = 48). The cut-off is $\lambda^2 = 0.3a^2$. The direction of cones indicates the direction in colour space.

Figure 2: Large lattice: Pseudoscalar density maxima, projected into 3 dimensions. Colour indicates position of maximum in fourth dimension, moving from red (z = 1) to green (z = 48). The cut-off is $\lambda^2 = 1.0a^2$. The direction of cones indicates the direction in colour space.

Figure 3: Electric field lines passing through a maximum of action. The colours are those of SU(2). The scale is such that close to the center the distance between 2 consecutive cones is half a lattice unit. Large lattice, $\lambda^2 = 1.0$.

Figure 4: Histograms for summed Wilson action $\Sigma(\lambda^2, R)$ (see eq. 22) for $\beta = 2.85$, size = $48^3 \times 64$ (full line), and $\beta = 2.70$, size = $32^3 \times 64$ (broken line).

Figure 5:  Small lattice ($32^3 \times 64$, $\beta = 2.70$): Pseudoscalar density maxima, projected into 3 dimensions. Colour indicates position of maximum in fourth dimension. The cut-off is $\lambda^2 = 1.0 \ a^2$.

Figure 6:  Large lattice: Pseudoscalar density $P^a(x)$ in a lattice plane running through a spike. The cut-off is $\lambda^2 = 0.15 a^2$. The direction of the cones is given by the orientation in colour space.

Figure 7:  Large lattice: Gauge fields of red colour in a lattice plane running through the spike shown in the previous figure. The cut-off is $\lambda^2 = 0.15 \ a^2$. The direction of the cones is given by 3 space-time components of the gauge fields.

Figure 8:  Large lattice: Electric fields around a spike. The cut-off is $\lambda^2 = 0.5 \ a^2$. The colours are those of SU(2).

Figure 9:  Electric fields around an artificial instanton with radius $\rho = 15 \ a$ in singular gauge. The cut-off is $\lambda^2 = 0.5 \ a^2$. The colours are those of SU(2).

Figure 10:  Large lattice: Electric field strengths of red colour in a lattice plane running through the spike shown in fig. 6. The cut-off is $\lambda^2 = 4.0 a^2$. The direction of the cones is given by the spatial orientation of the fields.

Figure 11:  Large lattice: Gauge fields of red colour in a lattice plane running through the spike shown in fig. 6. The cut-off is $\lambda^2 = 4.0 a^2$. The direction of the cones is given by three space-time directions of the gauge fields. The magnitude of the pseudoscalar density is indicated by golden circles.

Figure 12:   Large lattice: Gauge fields of green colour in a lattice plane running through the spike shown in fig. 6. The cut-off is $\lambda^2 = 4.0a^2$. The direction of the cones is given by three space-time directions of the gauge fields. The magnitude of the pseudoscalar density is indicated by golden circles.

Figure 13:   Large lattice: Gauge fields of blue colour in a lattice plane running through the spike shown in fig. 6. The cut-off is $\lambda^2 = 4.0a^2$. The direction of the cones is given by three space-time directions of the gauge fields. The magnitude of the pseudoscalar density is indicated by golden circles.

Figure 14:   Large lattice: Pseudoscalar density, projected into 3 dimensions. Colour indicates position in 4th dimension. 30 cooling steps have been executed with $c = 0$. The positions of the peaks are found numerically after cooling.

Figure 15:   Large lattice: Gauge fields of one SU(2) colour, at positions of spikes which are determined by Fourier filtering. Positions are projected into 3 dimensions, and colour indicates 4th dimension. The configuration has been cooled by 30 steps with $c = 0.$, after transforming it to Landau gauge.

Figure 16:   Large lattice: Gauge fields (3 components only) in a plane for SU(2), avoiding spikes. The cut-off is $\lambda^2 = 2.0a^2$.

Figure 17:   Gauge fields in a plane for non-compact QED. The cut-off is $\lambda^2 = 2.0a^2$, and the lattice size is $48^3 \times 64$.

Figure 18: Effective masses for longitudimal and transverse correlations of gauge fields. From top to bottom: Tranverse QED (open circles), transverse SU(2) (open squares), longitudinal QED (open diamonds), longitudinal SU(2) (full circles). The lattice size is $32^3 \times 64$, the Fourier filtering parameter $\lambda^2 = 1.0\ a^2$. For SU(2), $\beta = 2.7$.

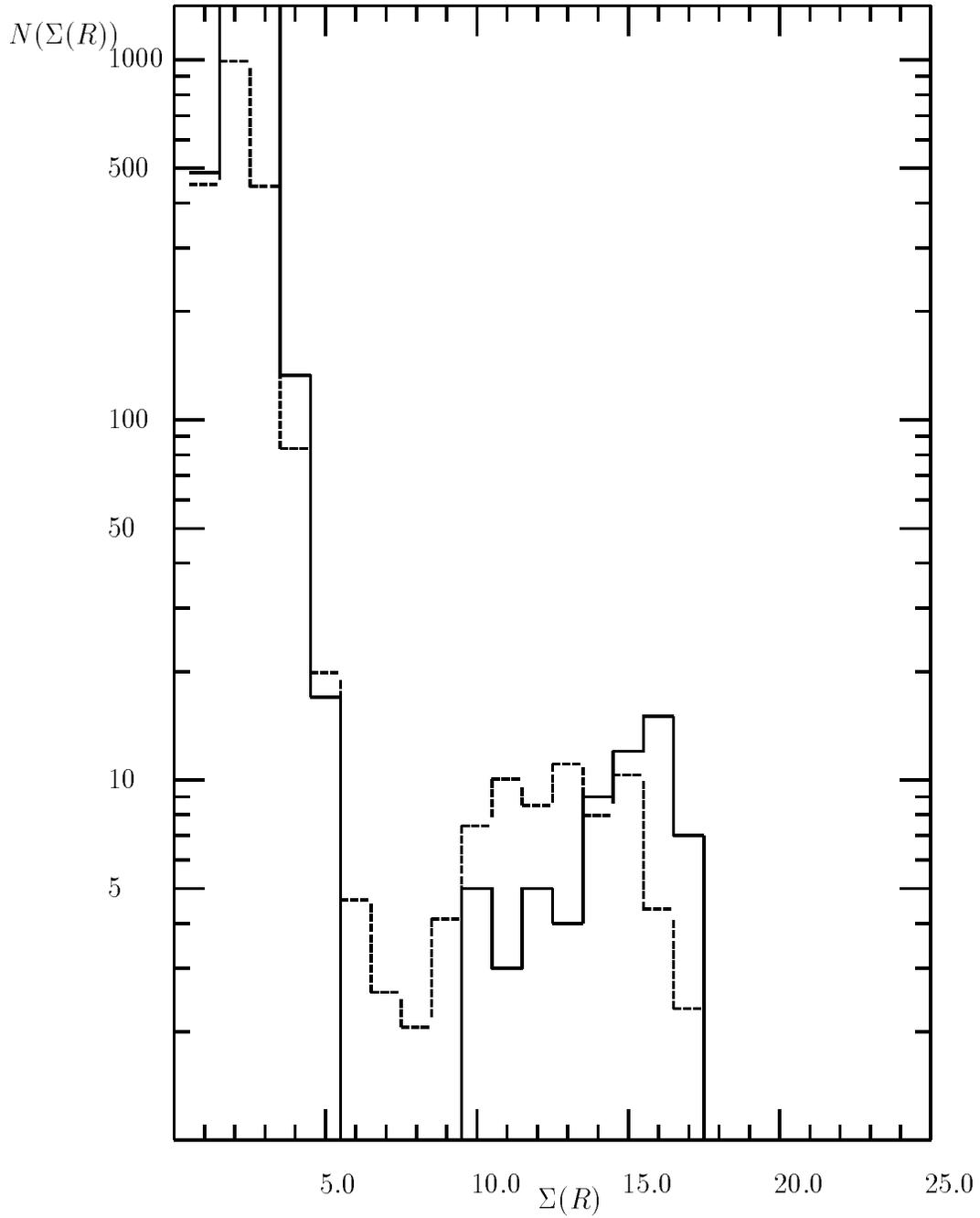

Figure 4: Histograms for summed Wilson action $\Sigma(\lambda^2, R)$ (see eq. 23) for $\beta = 2.85$, size $= 48^3 \times 64$ (full line), and $\beta = 2.70$, size $= 32^3 \times 64$ (broken line).

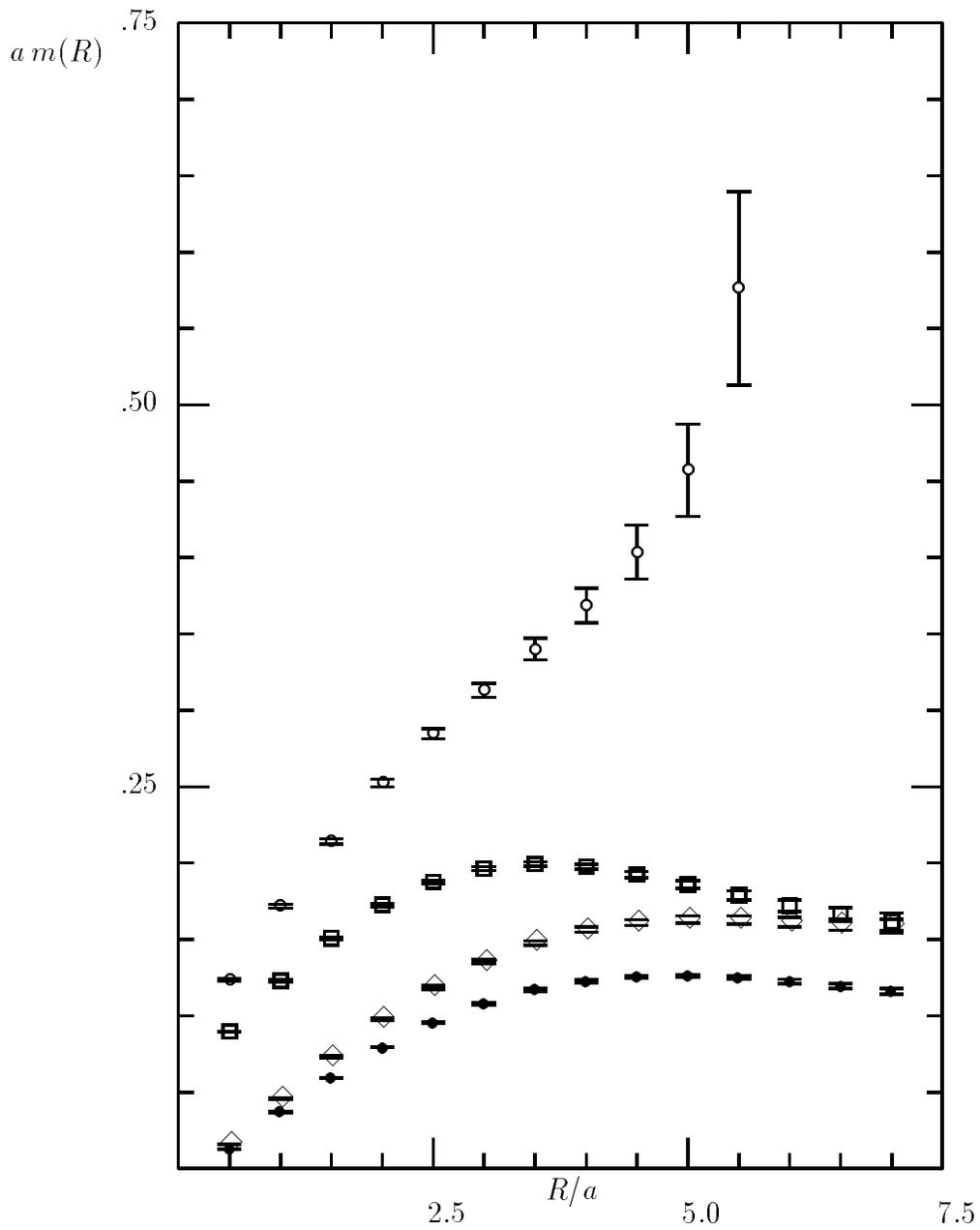

Figure 18: Effective masses for longitudimal and transverse correlations of gauge fields. From top to bottom: Tranverse QED (open circles), transverse SU(2) (open squares), longitudinal QED (open diamonds), longitudinal SU(2) (full circles). The lattice size is $32^3 \times 64$, the Fourier filtering parameter $\lambda^2 = 1.0\ a^2$. For SU(2), $\beta = 2.7$.